\documentstyle[multicol,aps,epsf,prl,amssymb]{revtex}
\begin{document}
\title{Variable-range-hopping in two-dimensional system.}

\author{D. N. Tsigankov, 
     and  A. L. Efros}
\address{ Department of Physics, University of Utah,
Salt Lake City, Utah 84112}

\maketitle
\begin{abstract}
Computer modeling of the VRH  conductivity in the
two-dimensional system has been done by kinetic Monte Carlo method,
which includes some new elements. Study of 
 the temperature dependence of the conductivity, testing  of the different 
 scaling relations, and study of the size effect
 show that the Efros-Shklovskii mechanism
of the VRH  is 
valid in the slightest details. It has been also shown that
simultaneous transitions of many electrons are not important. 
 The reasons of disagreement with previous works 
are thoroughly analyzed.
\end{abstract}


\begin{multicols}{2}

The concept of the variable range hopping (VRH) of localized electrons
 belongs to Sir Mott\cite{mott}. He has considered the phonon assisted
 tunneling of localized electrons between different sites. Mott has
 found that the typical hopping length for a two-dimensional (2D) case
 $R_M\approx(a/3) (T_M/T)^{1/3}$ increases with the decreasing
 temperature.  This is the origin of  the term  ``VRH''. Here
 $T_M=\beta_M(g_0 a^2)^{-1}$, $a$ is the localization length of an
 electron, $g_0$ is the density of states at the Fermi level, $\beta_M$
 is a numerical coefficient.  The conductivity obeys the Mott's
 law\cite{mott,pre}
\begin{equation}
\sigma_M \sim (\gamma e^2/T)\exp\left(-(T_M/T)^{1/3}\right),
\label{M}
\end{equation}
where  $\gamma$ is a typical phonon  frequency of the order of
$10^{12}{\rm sec^{-1}}$.

Electron-electron interaction  has added a new chapter to the
theory of VRH.  The study of the interaction in localized regime has been
initiated by Pollak\cite{Pollak70} and
 Srinivasan\cite{sri}. Later on Efros and Shklovskii\cite{ES} have
shown that the single particle density of states (DS) in 2D case
linearly tends to zero as energy tends to the Fermi energy. This
phenomenon, which is called the Coulomb gap(CG), is due to the long range
part of the Coulomb interaction which, in a sense, remains
non-screened. At large disorder the DS $G(\epsilon)=2|\epsilon|/\pi
e^4 $ has a {\em universal} form. The DS is given by the only
combination of the energy and electron charge which has a proper
dimensionality. In fact, the CG results from the Coulomb law
and from the discrete nature of electron charge.

  Simple arguments based upon single-electron excitations 
 show that the VRH with the Coulomb interaction
(CI) obeys Efros-Shklovskii (ES) law\cite{ES}
\begin{equation}
\sigma_c \sim (\gamma e^2/T)\exp\left(-(T_0/T)^{1/2}\right),
\label{ES}
\end{equation}
where $T_0=\beta_0 e^2/a$. The percolation approach (See references
in\cite{lev}) gives $\beta_M=13.8, \beta_0=6.5$. The hopping length
$R_C\approx (a/4)(T_0/T)^{1/2}$.

Many experimental works on the 2D structures report the ES law, but
very often the data are ambiguous because one should have a really
large interval of $\sigma$ to distinguish between the two laws.  We
think that the most important results have been obtained recently by
groups of Jiang and Dahm and Adkins\cite{dahm}. Using an artificial
screening provided by metallic electrode, parallel to a plane with the
2D electrons, they have proved that the VRH transport 
reflects the crucial feature of the CG, {\em the sensitivity
to the long range interaction}.

The theoretical understanding of the VRH is controversial. Efros and
Shklovskii argue that in 2D case only the single electron
transition with the typical hopping length are important\cite{ef,rev} and
the VRH has the same universal nature as the CG. Pollak and his
followers\cite{Pollak,Perez,mob,sch} claim that both sequential and
simultaneous transitions of different groups of electrons are very
important for the VRH, so that the physics of the CG which is
valid for the long hops only is not relevant. These claims are based
 upon the theoretical idea 
that simultaneous transition of many electrons may provide
 a gain in the activation\cite{Pollak}. Computer simulations of the VRH
 and the similar
phenomena made by this group also show importance of
the  many-electron transport 
 both in 3D and 2D cases but no  new $T$-dependence
 of the conductivity has been proposed.   The simulation of
 our group\cite{lev} reveals scaling
typical for the CG, but in this work  a simplified model has been used
which does not take into account any simultaneous transitions and
 underestimates sequential transitions with the short length.

We  present here  the results of computations which takes into
account all the sequential transitions and the most important
simultaneous transitions of two electrons. We show that  the VRH both
in the CI case and without interaction (M-case), closely
 obeys the laws Eqs. (\ref{ES}),(\ref{M}) and different
scaling laws which reflect the physics of the both cases. We also
confirm earlier prediction that simultaneous transitions of many
electrons are not important for the VRH in 2D case at low temperatures
and analyze the discrepancy with the
simulations by Pollak group.

We consider the standard lattice model with the Hamiltonian
\begin{equation} \label{Eq:Hamiltonian}
H = \sum_{i}\phi_{i} n_{i} + {1 \over 2} \sum_{i \neq j} {1 \over
    {r_{ij}}} \left(n_i - \nu \right) \left(n_j - \nu \right),
\label{ham}
\end{equation}
where $n_i = 0, 1$ are the occupation numbers.
  The quenched random site energies $\phi_i$ are distributed
uniformly within the interval $[-1, 1]$,  $\nu$ is the average
occupation number,which is taken to be 1/2 everywhere below. The
magnitude of the quenched disorder is enough to provide the universal
CG
 at all energies important in our temperature
range\cite{pik}. Here and below the lattice constant is a length unit
and the nearest Coulomb energy which is equal to the amplitude of 
the disorder is  both energy and temperature unit. In the M-case the 
interaction term in Eq. (\ref{ham}) is neglected.
\begin{figure}[b]
\centerline{\epsfxsize=3.4in\epsfbox{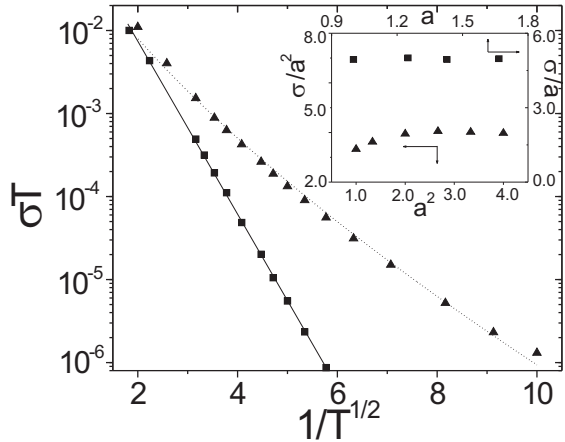}}
\caption{The temperature dependence of the conductivity as obtained by
simulations for the M-case ($\blacktriangle$) and for the CI-case
($\blacksquare$). Averaging over different disorder realizations is
performed to get the error bars equivalent to the sizes of the
symbols. Dashed line shows the fit by the Mott's law Eq. (\ref{M}),
while solid line represents ES-law Eq. (\ref{ES}). The localization
length $a=1$. The inset shows the test of the scaling laws
Eq. (\ref{scaling}). Here $Ta^2=0.04$ for the M-case ($\blacktriangle$)
and $Ta=0.1$ for the CI-case ($\blacksquare$) (See the explanations in
the text) }
\end{figure}

To simulate the VRH by the kinetic Monte Carlo (MC) method  we add an electric
 field $E$ and consider the periodic 
boundary conditions (See details in Ref\cite{lev}). To get conductivity
we calculate the dipole moment due  to electron transitions in the
 $E$-direction and divide it by the number of MC steps and by $E$.
This assumes that  our time unit is $1/\gamma$. We check that the result is
 $E$-independent. In fact we take $E$ from the 
condition $ER_{C,M}=0.1 T$.  The array size is taken $100\times 100$ for the temperatures above 0.05 and $200\times 200$ for the temperatures below 0.05.
 In the first approximation we neglect simultaneous transitions
of many electrons. The new element of our algorithm is
 that we do not
 include tunneling probability into MC process\cite{ofer}.
 Instead we choose  a  pair of sites $i$ and $j$
 with the probability $\exp(-2r_{ij}/a)/Z_1$, where $Z_1$ is the  normalization
factor. 
Then the transition is rejected if both sites are occupied or empty,
 and finally it is performed with the probability
 $1/(1+\exp(-\epsilon_{ij}/T))$, where $\epsilon_{ij}$ is the energy
 difference between the two configuration.  This new scheme dramatically
 decreases the runtime of the simulation and leads to $L^2$, rather
 than $L^4$, algorithm in number of MC steps. Note that for the
 CI-case each MC step consists of  $L^2$
 operation itself due to  recalculations of site energies after
 each transition.

To reveal and compare the physics of the VRH in both CI- and M-case we
perform both simulations.  Fig. 1 shows that the laws
Eqs. (\ref{ES}),(\ref{M}) are reproduced very good for both CI- and
M-cases. From solid and dashed line one gets $\beta_0=5.8$ and
$\beta_M=13.0$ respectively, that agrees well with the
percolation values cited above. Note that the temperature where two
curves become close to each other marks the upper temperature limit of
the VRH. At even higher temperatures the two curves coincide and the
conductivity decreases with the increasing temperature which manifests
transition to a diffusion regime.

To test  the origin  of the VRH we check the scaling
laws which reflect an intimate physics of both CI- and M-cases: $T_0$ and 
$T_M$ respectively are the only characteristic temperatures (energies) in the 
problem. These laws
can be written in the universal form
\begin{equation}
\begin{array}{c}
 \sigma_M = (\gamma e^2/T_M)f_M\left(T_M/T\right), \\
 \sigma_{CI} = (\gamma e^2/T_0)f_{CI}\left(T_0/T\right),
\end{array}
\label{scaling} 
\end{equation}
where $f_M(x)$ and $f_{CI}(x)$ are some functions. 
It follows from the scaling laws Eq. (\ref{scaling}) and the above
definitions of $T_M$ and $T_0$ that $\sigma_M/a^2$ and $\sigma_{CI}/a$
are functions of $Ta^2$ and $Ta$, respectively. Thus, if one changes
both $T$ and $a$ in the M-case in such a way that $Ta^2$ is constant
then $\sigma_M/a^2$ will be also constant. For CI-case one should
change $T$ and $a$ keeping $Ta$ constant to get $\sigma_{CI}/a$
constant. This test has been successfully performed and the convincing 
 results are shown in the inset of Fig.1. Note that the small
 deviation of $\sigma/a^2$ from the constant value for the M-case is
 due to the lattice effect. Since $R_M < R_C$, the lattice effect in
 the M-case appears at larger $a$ than in the CI-case.

In Fig.2 we present the results for the size effect in the CI-case. We
have calculated the average value of the conductivity
 {\bf av}$=\sigma_L/\sigma_{\infty}$, where $\sigma_L$ is the logarithmic
average of the conductivity over arrays with different disorder at
given $L$ and $T$, $\sigma_{\infty}$ is the conductivity at the maximum
$L$ at a given $T$. We have also calculated dispersion 
{\bf dis}$=\delta\sigma/\overline{\sigma}$, where $\delta\sigma$ is 
the dispersion of the conductivity and 
$\overline{\sigma}$ is the
arithmetic average of the conductivity  at given $L$ and $T$.

We did these computations  not only to get an idea how
 large the size of an array should be to obtain the  macroscopic value of
the  average conductivity and what value of the dispersion one should expect
 at a given $L$ and $T$. Another reason is that
  the $L$- and $T$-dependences of the average 
and the dispersion give us a new insight into  the physics of the VRH. 
One can see from Fig. 2 that the dispersion is a function of only
 one argument $TL$, as it follows from the ES theory. Indeed,
 the correlation length $L_C$  of the percolation network
is\cite{book} $L_C=R_C(T_0/T)^{\nu/2}$. Since the exponent of the 
correlation length  $\nu=1.33$, one gets that $L_C$ is very close to $1/T$.
Thus, $L/L_C\approx LT$ which explains the $L$- and $T$-dependence of 
the dispersion. 

The behavior of the average is completely different. It does not
change until very low values of $TL$. At even smaller $TL$ it
decreases dramatically and it is not any more a function of $TL$. We
believe that it  is a function of $L/R_{CI}\sim L\sqrt{T}$, and 
 the decrease of the average conductivity appears due to the hard gap 
in the DS. It is known\cite{bar} that the hard gap between occupied and empty 
states results from the size effect and its value is  of the order of $1/L$.
At $L<R_C$ the energy stripe of the VRH is completely within this gap.
We suggest that this is the reason of the decrease of the average conductivity.
Fig. 2 confirms qualitatively this suggestion. Unfortunately,
we are unable to check it quantitatively.
\begin{figure}[t]
\centerline{\epsfxsize=3.4in\epsfbox{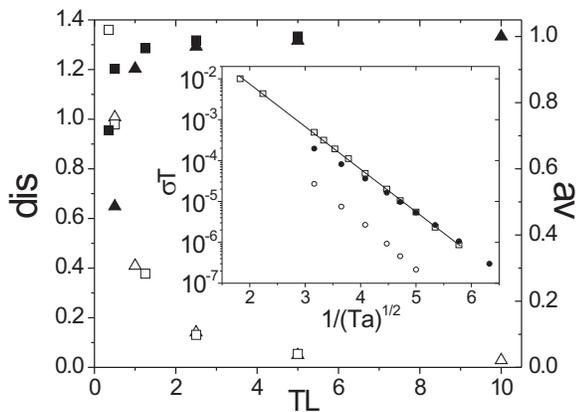}}
\caption{ The size dependence of the dispersion (left axis, opened symbols)
 and of the logarithmic average of the conductivity (right axis, solid symbols)
at $T=0.1$ ($\triangle, \blacktriangle$) and $T=0.05$
 ($\square, \blacksquare$) for the  CI-case. 
Inset: the  conductivity at the presence of the simultaneous transitions
of two electrons for the CI-case. The  contribution to the conductivity
 due to single
hops ($\bullet$) and due to double hops ($\circ$) at $a=0.5$
 should be compared. The 
 conductivity with single hops only  at $a=1$ ($\square$) 
is also shown.}
\end{figure}
Now we come to the  simultaneous transitions of many electrons in the
interacting system. We consider only two-electron transitions,
 which are the most important.

Suppose that in the initial state electrons occupy sites 1 and 3 and
in  the final
  state they are in the sites 2 and 4. The activation energy of this process
$\epsilon_{1234}$ is the energy difference between the final and
 the initial states. Due to Pollak it may be  smaller   than
 the sum of the energies of the single-electron transitions. Moreover,
the increase in the activation exponent  may  beat an extra tunneling
 exponent, which appears because of the tunneling of two electrons.
 That is why Pollak and his followers insist on many-electron processes.

First we calculate the probability of the process.  Let us consider
the case when $r_{12}, r_{34}\ll R$, where $R$ is the distance between
pairs (1,2) and (3,4).  Following Ref.\cite{Pollak} we concentrate on
the single phonon transitions.  It is obvious that these transitions
only cannot provide anything but independent hops of electrons. The
result is different if one considers the CI as well. We think that the
amplitude of  the most important 
process is described by the second order perturbation theory. It consists of
 a product of two matrix elements. The first one gives 
  the electron transition  in one of the close pairs, say (1,2), due to the 
phonon absorption  without 
 energy conservation. This element has a factor $\exp(-{r_{12}/a})$. The system comes to the intermediate state, where 
sites 2 and 3 are occupied. The second element describes the
 transition of the second electron from 3 to 4 in the external
 electrostatic potential created by the dipole (1,2). This element has a factor
$\exp(-r_{34}/a)/R^3$. The $R$-dependence is due to the
 dipole-dipole interaction. Use of the Golden rule gives the correct
energy  conservation law and a factor $\sim\exp(- 2(r_{12}+r_{34})/a)/R^6$.
The result is the same as for  the F\"{o}rster diffusion of molecular 
excitons \cite{Forster}, very similar but not identical problem. The transition
rate used in Ref.\cite{Perez,mob,sch} and many others is taken to be $R$-independent at large
 $R$. We cannot
understand this result because to get the common energy conservation law one
should assume that the remote pairs can ``talk'' to each other, which means 
there should be an interaction between them with some $R$-dependence. We
 argue below that the contradiction between our results originates from
the $R$-dependence.

 To accommodate this fast decay with $R$ we assume that the 
 simultaneous transitions occurs only  
if the initial or final position of the second electron is within the
spheres with the radii $R_{C}$ from the initial or final position of the
first electron. Note that this assumption overestimates the role of
 the double electron transitions  in the most important case when one electron 
makes a long hop $R_C$ and the other one makes a short hop. The interaction between such pairs decreases  (as $1/R^4$) at distances smaller than $R_C$.

To simulate the double hop transport we use the same idea as
 for single hop case. This time, however, we
choose a pair of sites $i$ and $j$ for single hop transition with the
probability $p_1= \exp(-2r_{ij}/a)/Z_2$,
and sites $i$, $j$, $k$ and $l$ for double hop with the probability 
$p_2=\exp(-2(r_{ij}+r_{kl})/a)/Z_2$, where $Z_2$ is found from the 
normalization condition for the sum of two probabilities.

 If the double hop is chosen, we check the occupation for both  pairs 
of sites $(i,j)$ and $(k,l)$ and then use
 the same Boltzman factor as before with $\epsilon_{ijkl}$ instead
 of $\epsilon_{ij}$.  The
 algorithm is also  $L^2$ in the number of MC steps. However, 
since 
the number of possible transitions of two electrons is much larger
 than the number of single transitions, one can show that there are
 $\approx 4R_{C}^2
 \exp (-2/a)$ attempts of double transition per one attempt of a
 single one, which makes simulation much more time consuming.

 The contribution of single and double hops to 
 the total conductivity can be separated by collecting their dipole
moments in the different cells.  We have checked  that the single
 hop contribution 
does not change in the presence of double hops.  The results of the
simulation  are shown on
the inset of Fig.2. We are able to calculate double hop contribution
for the case $a=0.5$, because at larger $a$ the value of $R_C$ is too large.
 The single hop contributions are shown for both $a=1,
0.5$. Note that   a small   deviation  from the universal behavior 
 at high temperatures is  a result  
of the lattice effect  for $a=0.5$. 
One can see  that the double hop contribution  is from
one to two orders of magnitude smaller than the single hop
contribution  at VRH region of temperatures. Moreover, the relative 
fraction  of double hops  in the total conductivity
decreases with decreasing  temperature. This result does not support
  Pollak's prediction\cite{Pollak} of
the gain in the activation energy for simultaneous transition of two
electrons as compared with the single electron transition. 

 The computational data, obtained in the papers
\cite{Perez,mob,sch} and many other works of this group  show the 
importance of the many-electron hops which  contradicts to
our result. In our opinion the $R$-independent probability of 
the simultaneous hops assumed by this group
 is responsible
 for the contradiction.

 Indeed, if  in a large array each pair has an equal
probability to make a transition with any other $L^2$ soft  pairs in the
system the number of transitions for the pair per time
is proportional to $L^2$. This is a giant overestimate of the role of
many-electron transitions. The abnormal size effect should be a
manifestation of this mistake.  Unfortunately, we have found only one
work\cite{mob}, where authors take care  about the size effect and this
work deals with the 3D case. However, we think that the infinite range 
interaction between pairs
should provide the same effect in the 3D-case. Looking at Fig. 3
 of Ref.\cite{mob} one can
see {\em an exponential} size effect without any signs of a saturation.
 The resistivity increases with the size
because the algorithm include the selection of states with the highest
transition rate among them. The selected states must consist of soft pairs
 which are far from each other and do not contribute to the dc conductivity.

Another reason for contradictions 
 is that  the method used
 in all these papers is applicable to the mesoscopic systems only.
It consists of  selecting  a number of low-energy states of the
total system to  study transitions between them. It looks nice
because it takes into account all  many electron transitions,
while we could afford the simultaneous transitions of two electrons
only. The problem is in  the necessary number of states, which
provides reasonable thermodynamic and kinetic description. This number 
 increases exponentially with the number of sites in the array.
Due to the sensitivity of the CG to the long range interaction 
 the size of the array $L$, which provides macroscopic regime,
increases  with decreasing $T$. One can estimate from Fig. 2 that
 $L\approx 3/T$ gives a reasonable dispersion.
 For example,  T=1/400 is the middle of
 the temperature interval in Ref.\cite{Perez}. It follows that $L$
 should be  at least 1000.
 Instead, authors compute
  few samples with different disorder at  $L\approx 22$.
 These samples have very
 different values of  conductivities and  different patterns of
the $T$-dependences (See Fig.1 of Ref.\cite{Perez}). This is a typical
 mesoscopic result.

Finally, our computer simulation  confirms   the
 ES theory of the VRH transport in 2D systems and shows that 
the attempts of its revision have no ground.
 
We are grateful to O. Biham and B. I. Shklovskii for the interest
 to this work and to 
E. I. Rashba for thorough discussion of the theory of simultaneous transitions.
The work have been funded  by the US-Israel Binational
Science Foundation  Grant 9800097, the computations have been made in CHPC of
the University of Utah.

\end{multicols}

\end{document}